\documentstyle[epsf,prd,aps,floats]{revtex}

\input{psfig.tex}
\tighten
\begin{document} 
\draft

\twocolumn[\hsize\textwidth\columnwidth\hsize\csname@twocolumnfalse\endcsname
\title{Inflation in Supersymmetric Cosmic String Theories}
\author{Anne-Christine Davis  \& Mahbub Majumdar}
\address{Department of Applied Mathematics \& Theoretical Physics (DAMTP),
University of Cambridge, Silver Street, Cambridge CB3 9EW, UK.  }
\maketitle
\begin{abstract}
We examine a non-Abelian SUSY $SU(2) \times U(1)$ gauge theory and a
SUSY $U(1)$ theory originally used to investigate the microphysics of
cosmic strings in supersymmetric theories. We show that both theories
automatically include hybrid inflation. In the latter theory we use a
$D$ term to break the symmetry.  SUSY is broken during inflation and
restored afterwards. Cosmic strings are formed at the end of
inflation. The temperature anisotropy is calculated and found to vary
as $(M_{GUT}/M_P)^2$.
\end{abstract}
\date{\today}
\pacs{}
]

\renewcommand{\thefootnote}{\arabic{footnote}}
\setcounter{footnote}{0}

\section{Introduction}

In recent years, supersymmetry (SUSY) has become increasingly favoured
as the theoretical structure underlying fundamental particle
interactions.  In this light it is natural to investigate possible
cosmological implications of SUSY.

Until recently, inflation and topological defects were thought of as
rival theories for the formation of density perturbations in the
Universe.  However, in supersymmetric theories it is natural for the
two to coexist.  In this paper we examine recently proposed
supersymmetric models which give rise to cosmic strings
\cite{nonabelian,abelian}. We show that these theories are
automatically hybrid inflation models. Previous work explored the
microphysics of the cosmic strings, but did not consider the resulting
cosmology. Whilst the cosmology of mixed models with cosmic strings
and inflation has previously been explored, the fact that
supersymmetric models with cosmic strings naturally inflate is not so
well known.

The theories we examine are based on both non-Abelian and Abelian
gauge symmetries, with symmetry breaking arising in two distinct ways.
The first model is a $SU(2) \times U(1) \rightarrow U(1) \times Z_2$
supersymmetric gauge theory with the gauge symmetry breaking
implemented with a $F$ term\cite{nonabelian}. The second theory is a
SUSY $U(1) \rightarrow Z_2$ gauge theory with a
Fayet-Illiopoulos $D$ term\cite{abelian}. In both cases it was
previously found that there were fermion zero modes and supersymmetry
was broken in the string core. The zero modes in the first theory
occurred in pairs, whilst there was a single zero mode in the second
theory. The effects of soft-SUSY breaking terms were investigated.  In
the former case: (1) SUSY breaking Higgs mass terms did not destroy
the zero modes; (2) SUSY breaking gaugino mass terms destroyed all
zero modes which involved gauginos; (3) SUSY violating trilinear terms
created Yukawa couplings which destroyed all zero modes.  In the
second theory, the zero mode survived soft-SUSY breaking terms. This
zero mode could move along the string resulting in the string carrying
a current. Such current-carrying strings could cause cosmological
problems \cite{vortons}.

The models we consider are particularly attractive for several
reasons.  (1) They incorporate both cosmic strings and hybrid
inflation, which has recently been shown to give good agreement with
observations \cite{joao,battye}.  (2) The inflaton field is typically much
smaller than $M_P$, allowing Planck era physics to be ignored. (3)
Inflaton physics can be taken to be normal GUT particle
physics. Normally the inflaton potential must suddenly ``turn off'' to
allow inflation to end.  For field values less than $M_P$, such
functions are often complicated. In hybrid inflation, the ending of
inflation is simply done by one of the other GUT fields. This
simplification allows many types of simple potentials to cause and end
inflation.  (4) By invoking supersymmetry, coupling constants often do
not need to be fine-tuned. \cite{linde,copeland,lyth}

In this paper we show that the two models we consider do indeed hybrid
inflate and lead to acceptable cosmic microwave
anisotropies. Supersymmetry is broken during inflation and a radiative
correction to the effective potential ensures that inflation ends.
SUSY is restored after cosmic strings are formed, though as previously
shown \cite{abelian}, it is broken in the string core. Inflation ends
before the GUT phase transition commences since the slow roll
condition is violated before the phase transition. Similar results
have appeared in different contexts
\cite{lyth,dvali1,dvali3,jean1,jean2}.

We proceed by reviewing the $SU(2) \times U(1)$ $F$-term SUSY gauge
theory and the $D$-term U(1) model.  Then we compute the radiative
correction to the effective potential for both models.  The CMB
anisotropy is then calculated. Finally we end with our conclusions.

\section{A $SU(2) \times U(1)$ F-term SUSY gauge theory}
             
The tree-level effective potential for a SUSY invariant theory is 

\begin{equation}
U = \sum_i |F_i|^2 + \frac{1}{2}
\sum_a D^a D_a
\label{susypot}
\end{equation}

\noindent
where the $F$-term (first term) is $F_i = \delta W[\phi_i] /\delta
\phi_i$, and $W[\phi_i]$ is the superpotential with the fermionic
terms set to zero.  The $D$ term (second term) is $D^a = -(\kappa +
g\sum_{ij} q_a \phi_i^{\dagger} (T^a)^i_j \phi^j $. Here $\kappa$ is a
Fayet-Illiopoulos constant and is nonzero if and only if the action
possesses a $U(1)$ symmetry. $T^a$ is the group generator and the sum
in (\ref{susypot}) is over all group generators. $g$ is the coupling
constant and $q_a$ is a dimensionless charge.   

The $SU(2) \times U(1) $ theory we are considering has the
superpotential\cite{nonabelian} 

\begin{equation}
W = \mu_1 S_0 ({\bf{\Phi}} \cdot {\bf{\tilde{\Phi}}} - \eta^2) + \mu_2 (S
{\bf{\Phi}}^T  \Lambda {\bf{\Phi}} + 
\tilde{S} {\bf{\tilde{\Phi}}}^T \Lambda {\bf{\tilde{\Phi}}})
\end{equation}

\noindent
where  ${\bf{\Phi}}$ and  ${\bf{\tilde{\Phi}}}$ transform as $SU(2)$
triplets, and $S_0, S, \tilde{S}$ are $U(1)$ scalars.  

The tree level effective potential is

\begin{eqnarray}
U & = & \mu_1^2 | {\bf{\phi}} \cdot {\bf{\tilde{\phi}}} -\eta^2|^2  + \mu_2^2
|2\phi_1 \phi_3 - \phi_2^2|^2  + \mu_2^2 | 2 \tilde{\phi_1}
\tilde{\phi_3} - \tilde{\phi_2}^2|^2 + \nonumber\\
  & {} &  |\mu_1 s_0 {\bf{\tilde{\phi}}} +2 \mu_2 s \Lambda \phi |^2 +   
 |\mu_1 s_0 {\bf{\phi}}   + 2 \mu_2 s \Lambda \tilde{\phi} |^2 + \nonumber\\
 &{}& e^2 | (\phi_1 +\phi_3) \phi^*_2 -
 (\tilde{\phi_1} + \tilde{\phi_3}) \tilde{\phi}_2^*|^2 + \nonumber\\
  & {} &  \frac{e^2}{2}
 (|\phi_1|^2 -|\phi_3|^2 -|\tilde{\phi_1}|^2 -|\tilde{\phi_3}|^2)^2 + \nonumber\\
  & {} &  \frac{e^2}{3} (|{\bf{\phi}}|^2 -|{\bf{\tilde{\phi}}}|^2 -2|s|^2
 -2|\tilde{s}|^2 )^2. \nonumber 
\end{eqnarray}

The minimum of the potential with regard to variations in the $SU(2)$
fields, $\phi, \tilde{\phi}$, leads to the constraint $\phi_1 =
\tilde{\phi_1}$ and $\phi_2 = \tilde{\phi_2} = \phi_3 = \tilde{\phi_3}
= 0$. The potential then becomes

\begin{equation}
U = \mu_1^2 | \phi_1^2 -\eta^2|^2  + 2 \mu_1^2
 |s_0|^2 | {\phi_1}|^2.
\end{equation}

\noindent
This possesses two minima.  A global minimum exists at $\phi_1 = \eta \equiv
M_{s_0}, s_0 = 0$ where supersymmetry is unbroken.  There is also a
SUSY violating local minimum at $\phi_1=0, s_0 > \eta \equiv
s_0^{crit}$.  This corresponds to the false vacuum.  We will assume a
chaotic inflation scenario, where $s_0$ is initially a random field
and $s_0 > s_0^{crit}$. Thus the $SU(2)$ fields are initially trapped
in a false vacuum and the potential  initially is

\begin{equation}
U = \mu_1^2 \eta^4.
\end{equation}

\noindent
The potential has no slope prompting $s_0$ to roll down to its
minimum.  However, eternal inflation is averted since SUSY is broken
and the tree level potential is no longer protected from radiative
corrections which cause inflation to end.

\section{A $U(1)$ $D$-term Gauge Theory} 

In this $U(1)$ theory we assume there exists a nonzero Fayet-Illiopoulos
constant, $\kappa$.  The superpotential is taken to be \cite{abelian} 

\begin{equation}
W = \alpha \Phi_0 \Phi_+ \Phi_-
\end{equation}

\noindent
where $\Phi_0, \Phi_+, \Phi_-$ have $U(1)$ charges 0, +1, -1,
respectively.  

If we redefine $D^a = -g(\kappa + \sum_{ij} q_a \phi_i^{\dagger}
(T^a)^i_j \phi^j) $, as is often done, the tree-level potential is \cite{lyth,jean2,dvali2}

\begin{eqnarray}
U & = &  \alpha^2|\phi_0|^2(|\phi_+|^2 + |\phi_-|^2) + \alpha^2 |\phi_+
\phi_-|^2 \nonumber\\
& {} & 
+ \frac{e^2}{2}(|\phi_+|^2 -|\phi_-|^2 + \kappa)^2.
\end{eqnarray}

\noindent 
$U$ possesses two minima -- a global minimum at $\phi_0 = \phi_+ =0,
\phi_- = \sqrt{\kappa}$ where supersymmetry is unbroken, and a SUSY
violating local minimum at $\phi_+ = \phi_- = 0, \phi_0 > (e/\alpha)
\sqrt{\kappa} = \phi_0^{crit}$. As before, we assume a chaotic
inflation scenario and that $\phi_0 > \phi_0^{crit}$. The fields are
in the false vacuum and the potential is

\begin{equation}
U = \frac{e^2}{2} \kappa^2.
\end{equation}

\noindent
As before this potential has no slope.  However, a SUSY breaking
induced radiative correction will add a slope.

\section{One Loop Radiative Corrections to the $SU(2) \times U(1)$ and $U(1)$
 Theories} \label{correction}

\subsection{ $SU(2) \times U(1)$ Theory}

We first consider the one loop radiative correction for the $SU(2)
\times U(1)$ theory after SUSY breaking.  

At one loop, $\phi$ and $\tilde{\phi}$ receive mass corrections.  The
$s, \tilde{s}, s_0$ fields will not acquire masses.  To see this,
shift the fields from their expectation values

\begin{eqnarray}
s   \rightarrow s' + s \,\,\,\,\,
& \tilde{s}  \rightarrow \tilde{s}' + \tilde{s} & 
\,\,\,\,\, s_0  \rightarrow s_0' + s_0 \nonumber\\
\phi_i \rightarrow  \phi_i' + \phi_i  & {} & 
\tilde{\phi}_i  \rightarrow  \tilde{\phi}_i' + \tilde{\phi}_i 
\end{eqnarray}

\noindent
where we take $s,\tilde{s},s_0,\phi_i,\tilde{\phi}_i$ to be the classical
vacuum expectation values. We assume that we are in the false vacuum
dominated case.  Thus

\begin{eqnarray}
s  =  \tilde{s} = \phi_i =  \tilde{\phi}_i = 0\,\,\, &{}&\,\,\,
s_0 \neq 0 
\end{eqnarray}

Upon substituting the shifted fields, and requiring $\phi_i' =
\tilde{\phi_i'}$ and  $s = \tilde{s}$, the $D$-terms drop out.  The
potential to leading order is 

\begin{equation}
U  =   \mu_1^2 | \vec{\phi}'^2 -\eta^2|^2  + 2 \mu_1^2  |s_0|^2 |
{\vec{\phi}}'|^2 + \dots
\end{equation}

\noindent
To find the masses of the fields substitute

\begin{equation}
\phi_i'  =  \frac{a_i + ib_i}{\sqrt{2}}.
\end{equation}

\noindent
This gives 

\begin{eqnarray}
U & = & 2 \cdot \frac{(\mu_1^2 |s_0|^2 - \mu_1^2 \eta^2)}{2} \sum_i
a_i^2 \nonumber \\
& {} & 2 \cdot \frac{( \mu_1^2 |s_0|^2 + \mu_1^2
\eta^2)}{2} \sum_j b_j^2 + {\rm non-mass\,\, terms}. \nonumber\\
\end{eqnarray}

Thus the $a_i$ acquire masses $ \mu_1^2 |s_0|^2 - \mu_1^2 \eta^2$ and
the $b_i$ acquire masses $\mu_1^2 |s_0|^2 + \mu_1^2 \eta^2 $. (There
are two sets of $a_i$ fields and two sets of $b_i$ -- hence the factor
of 2.) The $\vec{\phi}$ ($\vec{\tilde{\phi}}$) field splits up
into two fields \cite{dvali1,dvali2,lyth}, $\vec{\phi} +
\vec{\phi^*}$,($\vec{\tilde{\phi}} + \vec{\tilde{\phi^*}}$) and
$i(\vec{\phi} - \vec{\phi^*})$, ($i(\vec{\tilde{\phi}} -
\vec{\tilde{\phi^*}})$) with masses $ \mu_1 |s_0|^2 \pm \mu_1^2
\eta^2$.  We now calculate the masses of the fermion partners.

For a general superpotential 

\begin{equation}
W[\Phi]  = f_i \Phi_i + \frac{m_{ij}}{2} \Phi_i \Phi_j +
\frac{\lambda_{ijk}}{3} \Phi_i \Phi_j \Phi_k
\end{equation}

\noindent
the fermionic mass portion of the Lagrangian is

\begin{equation}
{\cal{L}}^F_m = -\frac{1}{2} ( m_{ij} + 2 \lambda_{ijk} \langle \phi_k
\rangle ) \psi_i
\psi_j  + {\rm H.C}
\end{equation}

In our case $m_{ij} = 0$ and the only non-zero scalar expectation
value is $s_0$.  Thus,

\begin{equation}
{\cal{L}}^F_m = -\mu_1 s_0 \psi_i
\tilde{\psi}_j  + {\rm H.C},
\end{equation}

\noindent
and the fermions have mass $\mu_1 s_0$.  

The 1-loop potential is found using the formula\cite{coleman} 

\begin{equation}
U(s_0)_{1-loop}  = \sum \frac{(-)^F}{64 \pi^2} M_i(s_0)^4 \ln
\frac{M_i(s_0)^2}{\Lambda^2}
\end{equation}

\noindent
where $\Lambda$ is an ultraviolet cutoff, and $F = \pm1$ for bosons/fermions.
Inserting the masses and adding in the zeroth order
part we have \cite{dvali1,lyth} 

\begin{eqnarray}
U_{eff}(s_0)  & = & \mu_1^2\eta^4(1 + 2 \cdot \frac{\mu_1^2}{64 \pi^2}(2
\ln(\frac{ \mu_1 |s_0|}{\Lambda})^2  \nonumber\\
&{}& 
+ (\frac{{s_0}^2}{\eta^2} - 1)^2 \ln (1 - \frac{\eta^2}{{s_0}^2}) 
+ (\frac{{s_0}^2}{\eta^2} + 1)^2 \ln (1 + \frac{\eta^2}{{s_0}^2})) \nonumber.
\end{eqnarray}

In the chaotic inflation limit, $s_0 \gg s_0^{crit} = \eta$, one finds

\begin{equation}
U_{eff}(s_0)  = \mu_1^2 \eta^4 (1 + \frac{\mu_1^2}{16 \pi^2}(\ln
(\frac{ \mu_1 |s_0|}{\Lambda})^2 + \frac{3}{2})). 
\end{equation}

\subsection{U(1) Theory with Fayet-Illiopoulos $D$-term}

The one loop correction for the $U(1)$ theory with a nonzero
Fayet-Illiopoulos constant, $\kappa$, is computed just as for the
$SU(2) \times U(1)$ theory. As before, the $\phi_+$ and $\phi_-$ split
up into scalars with masses $\alpha^2 |\phi_0|^2 \pm e^2 \kappa$.  The
one loop potential in the limit of a large inflaton field $\phi_0$ is
\cite{dvali2,lyth}

\begin{equation}
U_{eff}(\phi_0)  = \frac{e^2 \kappa^2}{2}(1 + \frac{e^2}{16 \pi^2}(\ln
(\frac{ \alpha |\phi_0|}{\Lambda})^2 + \frac{3}{2})). 
\end{equation}

\section{The End of Inflation}

Inflation occurs as long as the slow-roll parameters $\epsilon, \eta$
remain small, ($\epsilon, \eta << 1$), where \cite{copeland}

\begin{eqnarray}
\epsilon  =  \frac{M_P^2}{16 \pi} (\frac{U_{eff}'}{U_{eff}})^2  \,\,\,
&{}& \,\,\,
\eta  =   \frac{M_P^2}{8 \pi} \frac{U_{eff}''}{U_{eff}} 
\end{eqnarray}

Writing $s_0 = x s_0^{crit}$  and using

\begin{equation}
U'_{eff} = \frac{ \mu_1^4 M_{s_0}^3}{32 \pi^2} 4x((x^2-1)\ln(1-\frac{1}{x^2}) +
(x^2+1)\ln(1+\frac{1}{x^2})) 
\end{equation}

In the limit of large $x$, this is 

\begin{equation}
U'_{eff} \approx \frac{\mu_1^4 M_{s_0}^3}{8 \pi^2 } \frac{1}{x}.
\label{uprime_lim}
\end{equation}

\noindent
We find for the $SU(2) \times U(1)$ case \cite{dvali1}

\begin{eqnarray}
\epsilon & = &  (\frac{\mu_1^2 M_P}{8\pi^2 M_{s_0}})^2 \frac{x^2}{16 \pi}
((x^2-1) \ln(1 - \frac{1}{x^2}) + (x^2 + 1) \ln(1 + \frac{1}{x^2}))^2
\label{ep_limit} \nonumber\\
\eta & = &  (\frac{\mu_1 M_P}{4\pi M_{s_0}})^2 \frac{1}{4 \pi}
((3x^2-1) \ln(1 - \frac{1}{x^2}) + (3 x^2 + 1) \ln(1 + \frac{1}{x^2}))
\label{eta_lim} \nonumber
\end{eqnarray}

\noindent
where we approximated $U_{eff} = \mu_1^2 \eta^4$. In the limit of
large $x$

\begin{eqnarray}
\epsilon & = &  \frac{\mu_1^4}{1024 \pi^5}(\frac{M_P}{M_{s_0}})^2 \frac{1}{x^2}
\nonumber\\
\eta & = &  -\frac{\mu_1^2}{64 \pi^3} (\frac{M_P}{ M_{s_0}})^2 \frac{1}{x^2}.
\end{eqnarray}

For $x \gg 1$, both $\eta, \epsilon$ are small.  For $x \approx 1$, which
corresponds to the GUT phase transition, both $\epsilon,\eta$
diverge.  Hence inflation ends \cite{dvali1} before the phase
transition; $x>1$.

The same conclusion holds for the $U(1)$ theory with a
Fayet-Illiopoulos $D$-term.

\section{Predicted CMB Anisotropies of the $SU(2) \times U(1)$ and
$U(1)$ theories}

\subsection{$SU(2) \times U(1)$ theory}

The temperature quadrapole anisotropy due to inflation can be found to
be \cite{densitypert}

\begin{eqnarray}
(\frac{\Delta T}{T})_{infl}  & \approx & \sqrt{\frac{32 \pi}{45 M_P^6}} \frac{
U^{3/2}_{eff}}{U'_{eff}}|_{x_Q} \nonumber\\
& \approx & \sqrt{\frac{32 \pi}{45}} \frac{8 \pi^2}{\mu_1}
(\frac{M_{s_0}}{M_P})^3 x_Q  
\end{eqnarray}

\noindent
where we have used equation (\ref{uprime_lim}). This can be written in terms of, $N_Q$, the number of e-folds before the
end of inflation at which the Hubble radius crossed the de-Sitter
horizon. We have

\begin{eqnarray}
N_Q & = & -\frac{8 \pi}{M_P^2} \int_{Q}^{end}
\frac{U(s_0)}{U'(s_0)} ds_0 \nonumber \\
& \approx & \frac{32 \pi^3}{\mu_1^2}(\frac{M_{s_0}}{M_P})^2 (x^2_Q
-x^2_{end}) \nonumber\\
& \approx & \frac{32 \pi^3}{\mu_1^2}(\frac{M_{s_0}}{M_P})^2 x^2_Q
\label{efold}
\end{eqnarray}

\noindent
where we have used the approximation equation (\ref{uprime_lim}).

Hence

\begin{equation}
(\frac{\Delta T}{T})_{infl} \approx  \sqrt{\frac{8\pi}{45}} \sqrt{ 8 \pi N_Q} (\frac{M_{s_0}}{M_P})^2
\end{equation}

The SU(2) phase transition produces cosmic strings which contribute
to the CMB anisotropy.  The anisotropy due to strings can be
approximated as  \cite{shellard}

\begin{equation}
(\frac{\Delta T}{T})_{string} \approx 9 G\mu
\end{equation}

\noindent
where $\mu$ is the string mass per unit length.  We will approximate 

\begin{equation}
\mu \approx M_{s_0}^2 
\end{equation}

The combined temperature anisotropy is then \cite{jean1}

\begin{eqnarray}
(\frac{\Delta T}{T})_{tot} & \approx & \sqrt{ (\frac{\Delta T}{T})^2_{infl} + 
(\frac{\Delta T}{T})^2_{string}} \nonumber\\
& \approx & \sqrt{81 + \frac{8 \pi}{45} \cdot 8\pi   N_Q}
(\frac{M_{s_0}}{M_P})^2. 
\end{eqnarray}

We will assume that $N_Q \approx 60$. From the CMB temperature
anisotropy value of $10^{-5}$ we find $M_{s_0} \approx
7 \cdot 10^{15}$GeV.  We can estimate $\mu_1$ from equation (\ref{efold}).
If the Hubble radius crosses the de-Sitter horizon at $x_Q \approx 10$
we find $\mu_1 \approx 2 \cdot 10^{-2}$.

The spectral index is \cite{dvali1}

\begin{eqnarray}
n  & =  & 1 - 6 \epsilon + 2 \eta \nonumber\\  
  & \approx & 1 - \frac{1}{N_Q} = 0.98.
\end{eqnarray}

where we have used (\ref{ep_limit}) and the result $\mu_1 \ll 1$

\subsection{$U(1)$ Theory with Non-Zero Fayet-Illiopoulos Term}

Calculations of the CMB anisotropy for the $U(1)$ Fayet-Illiopoulos
case is identical to the $SU(2) \times U(1)$ case except  the
string tension, $\mu$, is identified as \cite{lyth}

\begin{equation}
\mu \approx M_{\phi_-}^2  \approx 2 \pi \kappa
\end{equation}

\noindent
since the phase transition occurs at $\langle \phi_- \rangle =
\sqrt{\kappa} = M_{\phi_-}$.

The CMB anisotropy then constrains $\kappa \approx 8 \cdot 10^{30}$ (GeV)$^2$.

\section{Comments and Conclusions}

We have examined two SUSY models which lead to strings and found that
that the models inflate -- via hybrid inflation.  The two models show
the different ways SUSY breaking can produce and end inflation -- by a
nonzero vev of a $F$ term, or a $D$ term. A $F$-term Abelian theory
would produce very similar results -- hence we have not included it in
the analysis.  We believe hybrid inflation to be a somewhat generic
feature of SUSY unified theories.  In fact, Jeannerot claims that
hybrid inflation is common to most unified theories which exclude
gravity with interactions of rank greater than or equal to five
\cite{jean2}.  Our results together with \cite{jean1} generalise this
and show that supersymmetric theories with cosmic strings in general,
are hybrid inflation models.

The CMB anisotropy of our models was found to be of order
$(M_{s_0}/M_P)^2$.  This is of the right magnitude, and allows GUT
scale physics to explain the CMB.

However, it is not clear how to extract the same inflation from a
higher energy theory such as supergravity or superstring theory.  $F$
term models in supergravity tend to cause all scalar fields to be
renormalised, even the inflaton.  This causes problems. It may
destabilise the flat directions and cause the slow-roll conditions to
be violated\cite{dvali2}. $D$ term supergravity models do not suffer
from this flaw.  But any theory relying on $D$ term inflation must not
be semi-simple. It's gauge group must contain a $U(1)$ factor.
Further, to preserve the flat directions of global SUSY, all fields
charged by the Fayet-Illiopoulos $U(1)$ must be driven to zero.  This
is not always possible.  Also, in the context of superstring
theory, if the $D$-term dominates any $F$-term during inflation a
runaway dilaton problem may occur. \cite{lyth}

Nevertheless, we believe that SUSY hybrid inflation is sufficiently
attractive to warrant further investigation.  We consider it
worthwhile to deduce whether hybrid SUSY inflation can easily fit
into a unified theory including gravity.

One caveat to our results is that the cosmic strings in the $D$-term
theory contain fermion zero modes that survive supersymmetry breaking.
Such strings could cause potential cosmological problems with the over
production of vortons\cite{vortons}.  Also, in the usual cosmic
string scenario, the Abelian Higgs model is used as a prototype
theory.  Recently this has been used in mixed string and inflation
scenarios \cite{joao,battye}.  However, the evolution of cosmic
strings in realistic theories, such as those considered in this paper,
is an open question and deserves investigation.
  
\section*{Acknowledgements}

This work was supported in part by PPARC.  M.M. is grateful to the
Cambridge Commonwealth Trust, Hughes Hall, and DAMTP for financial
support while this work was being completed.

\end{document}